\documentstyle[twoside,fleqn]{an-art}

\input plotting.sty

\begin{document}
\def\address #1END {{\vspace{9mm}\noindent\small Address of the author: \medskip \\ #1}}
\def\addresses #1END {{\vspace{9mm}\noindent\small Addresses of the authors: \medskip \\ #1}}

\begin{titlepage}

\setcounter{page}{1}

\def\makeheadline{\vbox to 0pt{\vskip -30pt\hbox to 50mm
{\small Astron. Nachr. 318 (1997) x, 1--x \hfill}}}
\makeheadline

\title {ROSAT public PSPC observations in the Las Campanas Redshift Survey}

\author{{\sc D. L. Tucker}, Batavia, Illinois, USA \\
\medskip
{\small Fermilab} \\
\bigskip
{\sc G. Hasinger}, Potsdam, Germany \\
\medskip
{\small Astrophysikalisches Institut Potsdam} \\
\bigskip
{\sc H. Lin}, Toronto, Ontario, Canada \\
\medskip
{\small Dept. of Astronomy, University of Toronto}\\
\bigskip
}

\date{Received ; accepted } 
\maketitle

\summary
The Las Campanas Redshift Survey, an optically selected survey which
contains 26\,418 galaxy redshifts, has been correlated with ``The
First ROSAT Source Catalogue of Pointed Observations with the PSPC,''
which contains 50\,408 sources from 2876 ROSAT pointed observations.
Ten matches were found.  The optical spectra of most of the ten
matches show weak narrow emission lines.  Due to their high x-ray
luminosities, their high x-ray--to--optical flux ratios, and the
evidence of rapid x-ray variability in the two brightest matches, we
interpret the majority of these objects to be narrow-line Seyfert
galaxies or ``hidden'' active galactic nuclei.  Of the ten matches,
only one galaxy shows the characteristics of a bona fide starburst.END

\keyw
catalogs -- surveys -- galaxies:  active -- galaxies:  starburst -- x-rays:  galaxiesEND
\AAAcla
???END
\end{titlepage}


\kap{Introduction}
Great strides have been made in recent years in resolving the soft
(0.5 -- 2~keV) extragalactic x-ray background (XRB) into discrete
sources (for a review, see Hasinger 1996).  These sources include
stars, galaxies, quasars and other active galactic nuclei (AGNs), and
groups and clusters of galaxies.  Most recent surveys of the soft XRB
have focused on finding the optical counterparts of x-ray sources from
flux-limited x-ray catalogues.  In this paper, however, we reverse the
process: we take an optically selected galaxy redshift survey -- the
Las Campanas Redshift Survey (LCRS; Shectman et al.\ 1996) -- and
search for x-ray sources within it, taking for our x-ray sample ``The
First ROSAT Source Catalogue of Pointed Observations with the PSPC''
(ROSATSRC; Zimmermann 1994).  That our approach is complementary to
the standard strategy is evident, since it is based upon a
magnitude-limited optical galaxy survey which therefore contains much
fainter x-ray sources (down to $\sim 3 \times
10^{-15}$~erg~s$^{-1}$~cm$^{-2}$) than most present-day x-ray surveys.
With such an approach, we can learn about the x-ray properties of a
typical, optically selected catalogue of galaxies.

\kap{The LCRS and the ROSATSRC}

The LCRS is an optically selected galaxy redshift survey which extends
to a redshift of 0.2 and which is composed of a total of 6 alternating
$1\fdg5 \times 80^{\circ}$ slices in the North and South Galactic
Caps.  Accurate $R$-band photometry and sky positions for program
objects have been extracted from CCD drift scans obtained on the Las
Campanas Swope 1-m telescope; spectroscopy has been performed at the
Las Campanas Du Pont 2.5-m telescope, originally via a 50-fiber
Multi-Object Spectrograph (MOS), and later via a 112-fiber MOS.  For
observing efficiency, all the fibers are used, but each MOS field is
observed only once.  Hence, the LCRS is a collection of 50-fiber
fields (with nominal apparent magnitude limits of $16 < m_R < 17.3$)
and 112-fiber fields (with nominal apparent magnitude limits of $15 <
m_R < 17.7$).  Recently completed, the LCRS contains 26\,418 galaxy
redshifts; in this paper, we consider only those 25\,327 LCRS galaxy
redshifts which lie in the 6 published slices (Shectman et al.\ 1996).

Our x-ray sample, ROSATSRC, contains 50\,408 x-ray sources in 2876
2$^{\circ}$ fields of the ROSAT Position Sensitive Proportional
Counter (PSPC; Pfeffermann et al.\ 1986).  This catalogue is composed
of PSPC fields in the public ROSAT Data Archive observed before June
1993.  The sources in the catalogue were extracted at a detection
threshold of likelihood $\ln P \ge 10$, which corresponds to a rate of
accidental detections of about 1\%.

\kap{Correlation of LCRS Galaxies with ROSATSRC Sources}

The typical errors in the sky positions for LCRS galaxies and ROSATSRC
sources are, respectively, $\sim$ 1 arcsec and $\sim$ 10 -- 20 arcsec.
Hence, to match LCRS galaxies with ROSATSRC sources, we follow a
three-step procedure:

\begin{enumerate}
\item{Within a search radius of 30 arcsec, match an LCRS galaxy 
with the closest ROSATSRC source (if any).  At this stage, we had
19 potential LCRS-ROSATSRC matches.}
\item{Cull matches which lie outside a 2-sigma positional error box.
This step removed 3 ``matches'' from the sample, leaving 16.}
\item{By visual inspection of the Palomar Digitized Sky Survey, remove
ambiguous identifications and sources which appear to be groups or
clusters.  This final step left a cleaned sample of 10 LCRS-ROSATSRC
matches.}
\end{enumerate}

Such a small number of matches is perhaps not too surprising: although
the LCRS covers 700~sq~deg of sky, the public ROSATSRC pointings only
overlap $\approx 9\%$ of this area ($\approx 60$~sq~deg)
non-redundantly; if one considers only the more sensitive inner 20
arcmin radius of the ROSATSRC fields, this fraction drops to $\approx
0.7\%$ ($\approx 5$~sq~deg).  The sky and (redshift) space
distributions of the LCRS galaxies and of the 10 LCRS-ROSATSRC matches
are presented in Figs.~1 \& 2.  The optical spectra of the
LCRS-ROSATSRC matches can be found in Fig.~3.

\kap{LCRS-ROSATSRC Sample Characteristics}

Since the LCRS is a {\em galaxy\/} survey (and thus selects {\em
against\/} objects which appear {\em stellar\/}), none of the
LCRS-ROSATSRC matches were quasars (which, nevertheless, are
relatively common in typical, x-ray selected surveys).  Indeed, we
find instead the following general characteristics for the
LCRS-ROSATSRC matches (see Table~1):
\begin{enumerate}
\item{Most have relatively narrow [OII]~3727, H$\beta$, and [OIII]~5007 
emission lines (FWHM $<$ 1000~km~s$^{-1}$).}
\item{In general, their soft (0.5 -- 2~keV) x-ray flux $f_X$ 
lies in the range $f_X \sim 0.1$ -- $1~f_R$, where $f_R$ is the
$R$-band (``optical'') flux.  [The x-ray fluxes $f_X$ were estimated
from the count rates assuming Galactic values of $N_{\rm H}$ and a power law
spectrum with energy index 1; the flux ratios were calculated
according to the formula
\begin{eqnarray*}
\log (f_X/f_R) & = & \log [f_X(\mbox{0.5 -- 2 keV})] + 0.4m_R + 5.7, 
\end{eqnarray*}
which is appropriate for the LCRS $R$-filter.]}
\item{They generally have soft (0.5 -- 2~keV) x-ray luminosities in the 
range $L_X \sim 10^{42}$ -- $10^{44}$~erg~s$^{-1}$  ($H_0 =
50$~km~s$^{-1}$~Mpc$^{-1}$, $q_0 = 0.5$).} 
\item{Their x-ray spectra are of moderate ``hardness,'' ranging typically 
between values of -0.5 and +0.5 for both the $H~R~1$ and $H~R~2$
hardness ratios.}  
\end{enumerate}
As such, their general properties closely resemble those of other
samples of x-ray luminous, narrow emission-line galaxies (e.g., Boyle
et al.\ 1995a,b; Carballo et al.\ 1995; Griffiths et al.\ 1996;
Romero-Colmenero et al.\ 1996).  

Furthermore, we have produced x-ray light curves for two of the
brightest and longest exposed LCRS-ROSATSRC galaxies, one of which
shows a factor of two increase in brightness on a timescale on the
order of a few hours and the other of which shows a
less-significant-but-suggestive variation on a similar timescale
(Fig.~4).

In summary, due to their relatively high luminosities and $f_X/f_R$
ratios and due to the evidence for rapid x-ray variability, we
interpret the majority of these objects as narrow-line Seyfert
galaxies or ``hidden'' AGN; of these galaxies, only one
(LCRS~B120117.2-032552) shows definite characteristics of a starburst.

\begin{acknowledgments}
DLT wishes to thank AIP researchers S. Allam, Dr. R. Assendorp,
Dr. Th. Boller, and Prof. K.-H. Schmidt for helpful discussions during
the course of this work.  The Las Campanas Redshift Survey has been
supported by NSF grants AST~87-17207, AST~89-21326, and AST~92-20460.
\end{acknowledgments}

\refer
\aba
\rf{Boyle, B.J., McMahon, R.G., Wilkes, B.J., Elvis, M.: 1995a, 
MNRAS 272, 462}
\rf{Boyle, B.J., McMahon, R.G., Wilkes, B.J., Elvis, M.: 1995b, 
MNRAS 276, 315}
\rf{Carballo, R., Warwick, R.S., Barcons, X., et al.: 1995, 
MNRAS 277, 1312}
\rf{Griffiths, R.E., Della~Ceca, R., Georgantopolos, I., et al.: 
1996, MNRAS 281, 71}
\rf{Hasinger, G. 1996: in {\em R\"ontgenstrahlung in the Universe}, 
eds. H.-U. Zimmermann, J.E. Tr\"umper, H. Yorke, MPE Report 263, p. 291}
\rf{Pfeffermann, E., et al.: 1986, Proc. SPIE 733, 519}
\rf{Romero-Colmenero, E., Branduardi-Raymont, G., Carrera, F.J., et al.: 
1996, MNRAS 282, 94}
\rf{Shectman, S.A., Landy, S.D., Oemler, A., et al.: 1996, ApJ 470, 172}
\rf{Zimmermann, H.-U.: 1994, IAU Circ. \#6102}
\abe

\addresses
Douglas L. Tucker\\
Fermilab\\
MS 127\\
P.O. Box 500\\
Batavia, IL 60510\\
USA\\
\\
G\"unther Hasinger\\
Astrophysikalisches Institut Potsdam\\
An der Sternwarte 16\\
D-14482 Potsdam\\
Germany\\
\\
Huan Lin\\
Dept. of Astronomy\\
University of Toronto\\
60 St. George St.\\
Toronto, Ontario M5S~3H8\\
Canada
END

\newpage

\bigskip
\begin{quote}
\aba
\begin{table}
\begin{flushleft}
\begin{tabular}{lrllrrrlll@{ }r@{ }r@{ }}
\multicolumn{12}{l}{Table 1} \\
\multicolumn{12}{l}{Observed characteristics of the LCRS-ROSATSRC sample.} \\
\noalign{\medskip}
\noalign{\smallskip}
\hline
\noalign{\smallskip}
Name$^{\rm a}$ &    $cz_{\sun}$$^{\rm b}$ & $m_R$$^{\rm c}$ & $M_R$$^{\rm d}$ & \multicolumn{3}{c}{$W_{\lambda}$$^{\rm e}$}
&  count rate$^{\rm f}$  &  $f_X$$^{\rm g}$  &  $\log L_X$$^{\rm h}$ &  $HR1$$^{\rm i}$  &  $HR2$$^{\rm j}$  \\
\cline{5-7}
        &       &       &       & [OII] & H$\beta$ & [OIII]
        &       &       &       &     \\
LCRS B  &       &       &       &  3727 & 4861 & 5007
        &       &       &       &     \\
\noalign{\smallskip}
\hline
\noalign{\smallskip}
002810.0-422515 & 54034 & 17.06 & -23.38   &   5.63  &   2.56 & $<2.$   & $82.7 \pm 7.5$  & 60.8  & 43.963 & $-0.31 \pm 0.07$ & $-0.44 \pm 0.11$ \\
020707.7-392308 & 29491 & 17.51 & -21.50   &   3.38  & $<2.$  &   8.14  & $8.72$          &  5.12 & 42.347 &  $0.00 \pm 0.00$ &  $0.00 \pm 0.00$ \\
033922.7-442138 & 22883 & 16.62 & -21.80   &  16.72  &   3.99 &   2.83  & $5.20 \pm 0.55$ &  3.14 & 41.910 &  $0.13 \pm 0.14$ & $-0.27 \pm 0.12$ \\
102524.8-022953 & 53070 & 17.48 & -22.91   & $<2.$   & $<2.$  & $<2.$   & $10.8$          & 10.6  & 43.190 &  $0.00 \pm 0.00$ &  $0.00 \pm 0.00$ \\
120011.5-033039 & 19341 & 15.71 & -22.33   &  16.28  & $<2.$  &   9.54  & $25.7 \pm 1.8$  & 19.1  & 42.546 &  $0.93 \pm 0.07$ &  $0.36 \pm 0.07$ \\
120117.2-032552 &  3925 & 15.73 & -18.77   &  89.02  &  41.60 & 123.72  & $0.45 \pm 0.15$ &  0.35 & 39.42  &  $0.94 \pm 0.82$ &  $0.28 \pm 0.34$ \\
120154.1-030604 & 47819 & 17.81 & -22.33   &  10.84  & $<2.$  &   4.56  & $1.27 \pm 0.36$ &  0.90 & 42.02  &  $0.12 \pm 0.25$ & $-0.05 \pm 0.24$ \\
124520.5-060334 & 29190 & 17.25 & -21.73   &   8.68  &   3.60 &   4.81  & $28.5 \pm 4.7$  & 19.7  & 42.922 & $-0.25 \pm 0.16$ & $-0.25 \pm 0.48$ \\
125310.5-051811 & 25960 & 17.07 & -21.64   &   2.18  & $<2.$  &   3.68  & $3.58 \pm 0.45$ &  2.55 & 41.932 &  $0.43 \pm 0.15$ &  $0.12 \pm 0.13$ \\
215759.7-413356 & 41143 & 17.28 & -22.50   &   3.83  & $<2.$  & $<2.$   & $33.5 \pm 3.10$ & 19.8  & 43.231 & $-0.01 \pm 0.09$ & $-0.22 \pm 0.11$ \\
%
%
\noalign{\smallskip}
\hline
\noalign{\smallskip}
\multicolumn{12}{l}{$^{\rm a}$ A description of the IAU-registered naming convention for individual LCRS galaxies, LCRS BHHMMSS.s-DDMMSS, can } \\
\multicolumn{12}{l}{be can found in the online ``Dictionary of Nomenclature of Celestial Objects,'' at {\tt http://astro.u-strasbg.fr/cgi-bin/Dic}.} \\
\multicolumn{12}{l}{$^{\rm b}$ Heliocentric recessional velocity in km s$^{-1}$.} \\
\multicolumn{12}{l}{$^{\rm c}$ LCRS $R$-band apparent magnitude.} \\
\multicolumn{12}{l}{$^{\rm d}$ LCRS $R$-band absolute magnitude ($H_0 = 50$~km~s$^{-1}$~Mpc$^{-1}$, $q_0 = 0.5$).} \\
\multicolumn{12}{l}{$^{\rm e}$ Rest-frame equivalent width (estimated by a Gaussian fit to the line profile) in \AA; the detection limit is $\approx$ 2.0\AA.} \\
\multicolumn{12}{l}{$^{\rm f}$ Source counts per $10^3$ seconds (vignetting corrected).} \\
\multicolumn{12}{l}{$^{\rm g}$ X-ray flux ($0.5 - 2.0$ keV) in units of $10^{-14}$ erg s$^{-1}$ cm$^{-2}$; assumes Galactic values of $N_{\rm H}$ and a power law spectrum with } \\
\multicolumn{12}{l}{energy index 1.} \\
\multicolumn{12}{l}{$^{\rm h}$ Log of the x-ray luminosity ($0.5 - 2.0$ keV) in units of erg s$^{-1}$ ($H_0 = 50$~km~s$^{-1}$~Mpc$^{-1}$, $q_0 = 0.5$).} \\
\multicolumn{12}{l}{$^{\rm i}$ $HR1 \equiv (H - S) / (H + S)$, where $H$ is the flux in the 0.4 -- 2.4 KeV band and $S$ is the flux in the 0.1 -- 0.4 KeV band.} \\
\multicolumn{12}{l}{$^{\rm j}$ $HR2 \equiv (H2 - H1) / (H2 + H1)$, where $H1$ is the flux in the 0.5 -- 0.9 KeV band and $H2$ is the flux in the 0.9 -- 2.0 KeV } \\
\multicolumn{12}{l}{band.}
\end{tabular}
\end{flushleft}
\end{table}
\abe
\end{quote}
\bigskip

\newpage

\begin{figure}
\epsfxsize=16.0truecm \epsfbox[-50 100 546 742]{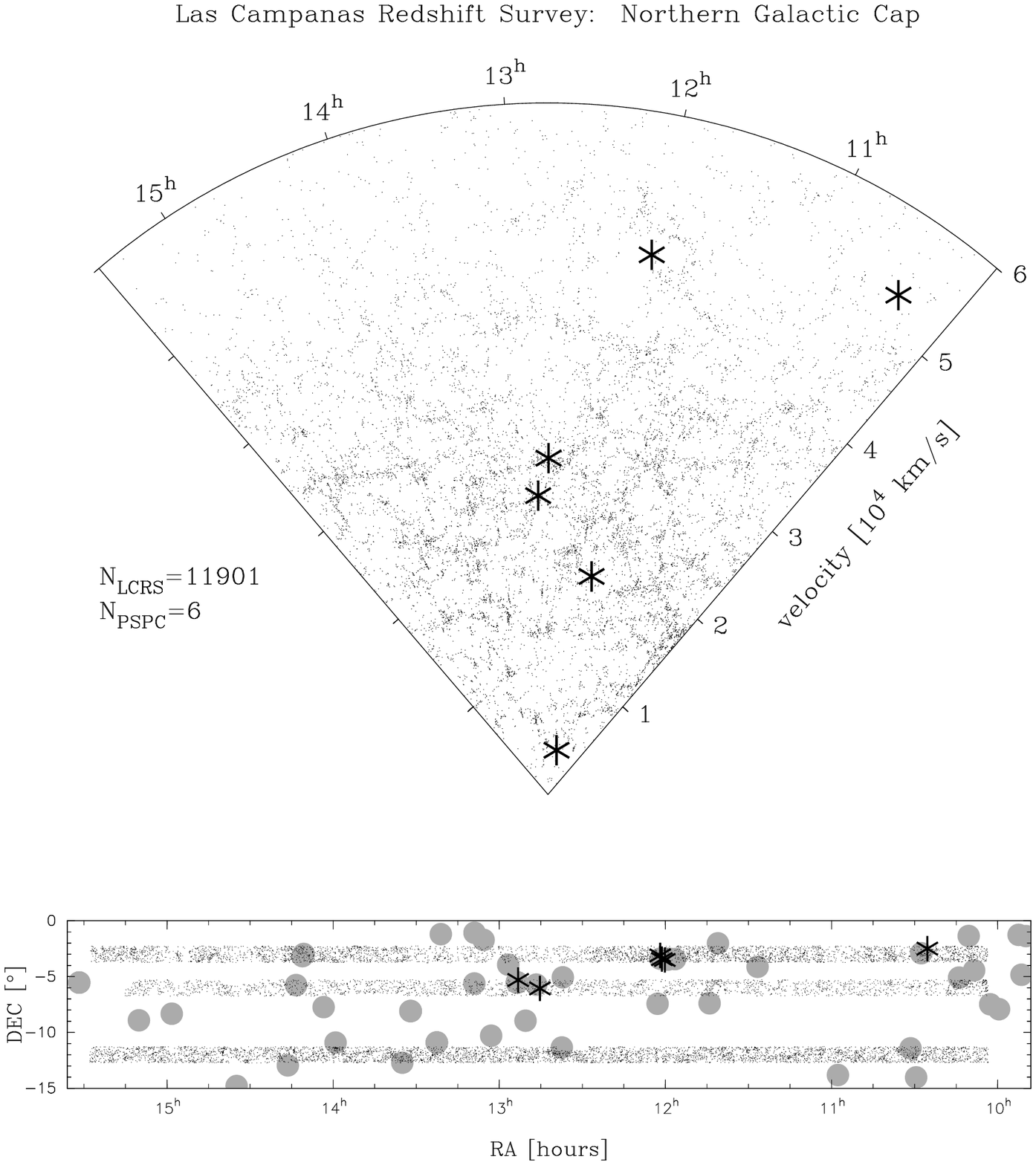} 
\vspace{-10mm}
\aba Fig.~1.  RA-velocity ``wedge'' map (top) and RA-DEC sky map
(bottom) for the three slices of the LCRS Northern Galactic Cap
sample.  In both maps, points denote LCRS galaxies and asterisks the
LCRS-ROSATSRC galaxy matches.  In the sky map, lightly shaded circles
represent the 2$^{\circ}$-diameter public PSPC fields near the region
surveyed by the LCRS.  Sky coordinates are Epoch 1950.0; radial
velocities are heliocentric.  \abe
\end{figure}

\newpage

\begin{figure}
\epsfxsize=17.0truecm \epsfbox[-50 100 546 742]{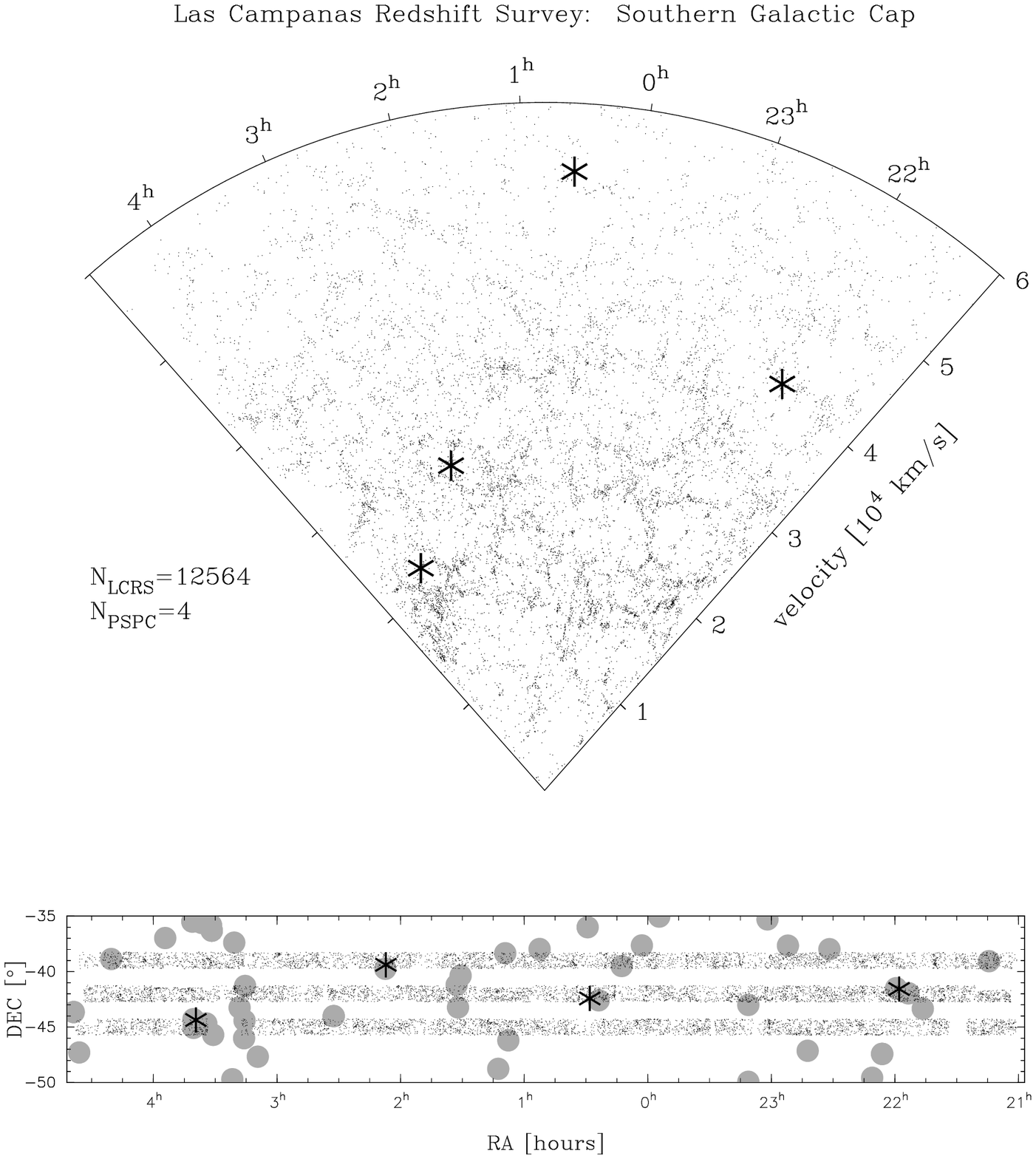} 
\vspace{-10mm}
\aba
Fig.~2.  RA-velocity ``wedge'' map (top) and RA-DEC sky 
map (bottom) for the three slices of the LCRS Southern Galactic Cap
sample;  symbols are as in Fig.~1.
\abe
\end{figure}

\newpage

\begin{figure}
\epsfysize=16.0truecm \epsfbox[20 99 491 593]{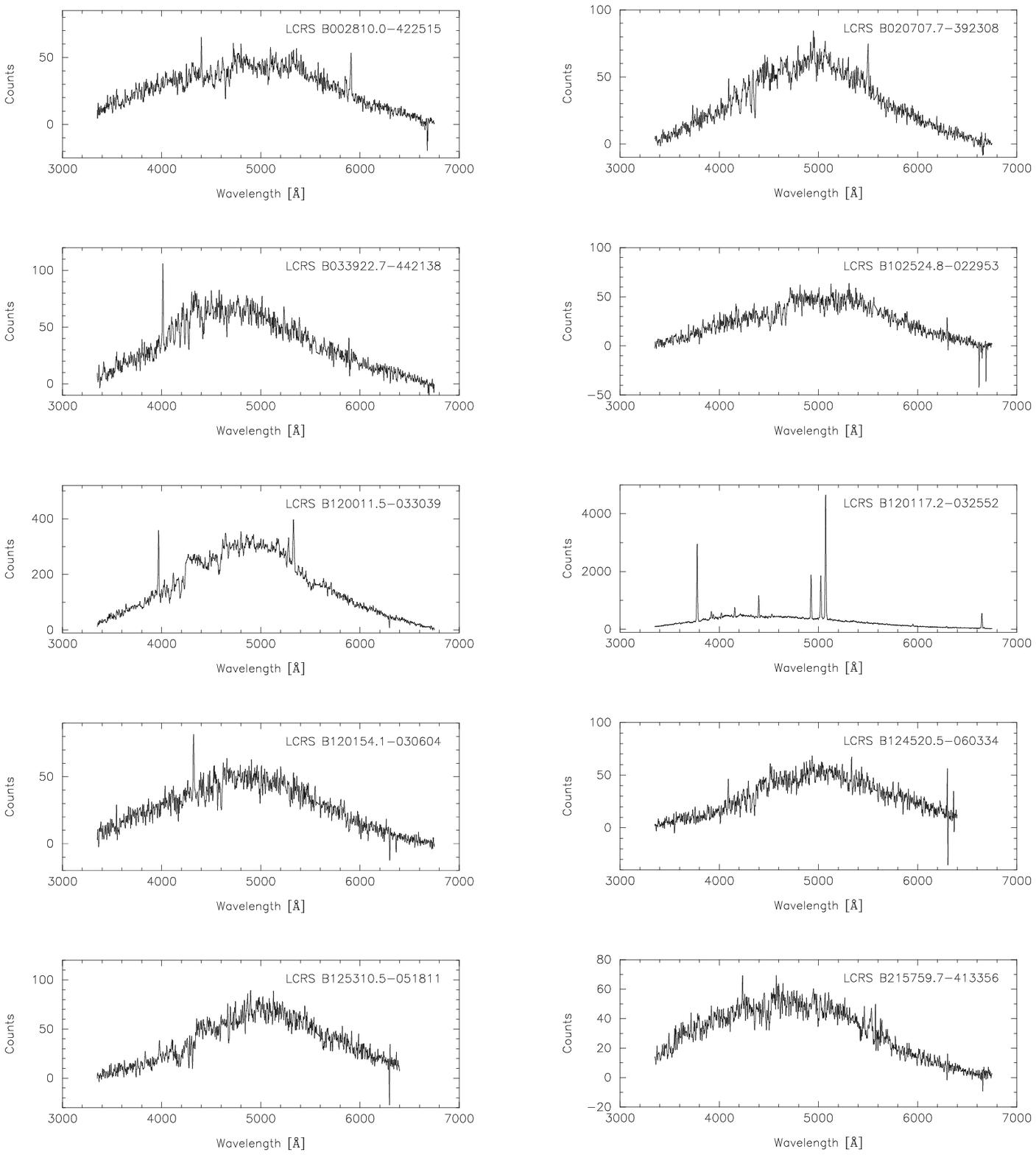}
\vspace{10mm}
\aba
Fig.~3.  Spectra of the matched LCRS-ROSATSRC galaxies (NOT flux calibrated).
\abe
\end{figure}

\newpage

\begin{figure}
\epsfxsize=16.0truecm \epsfbox[75 300 487 600]{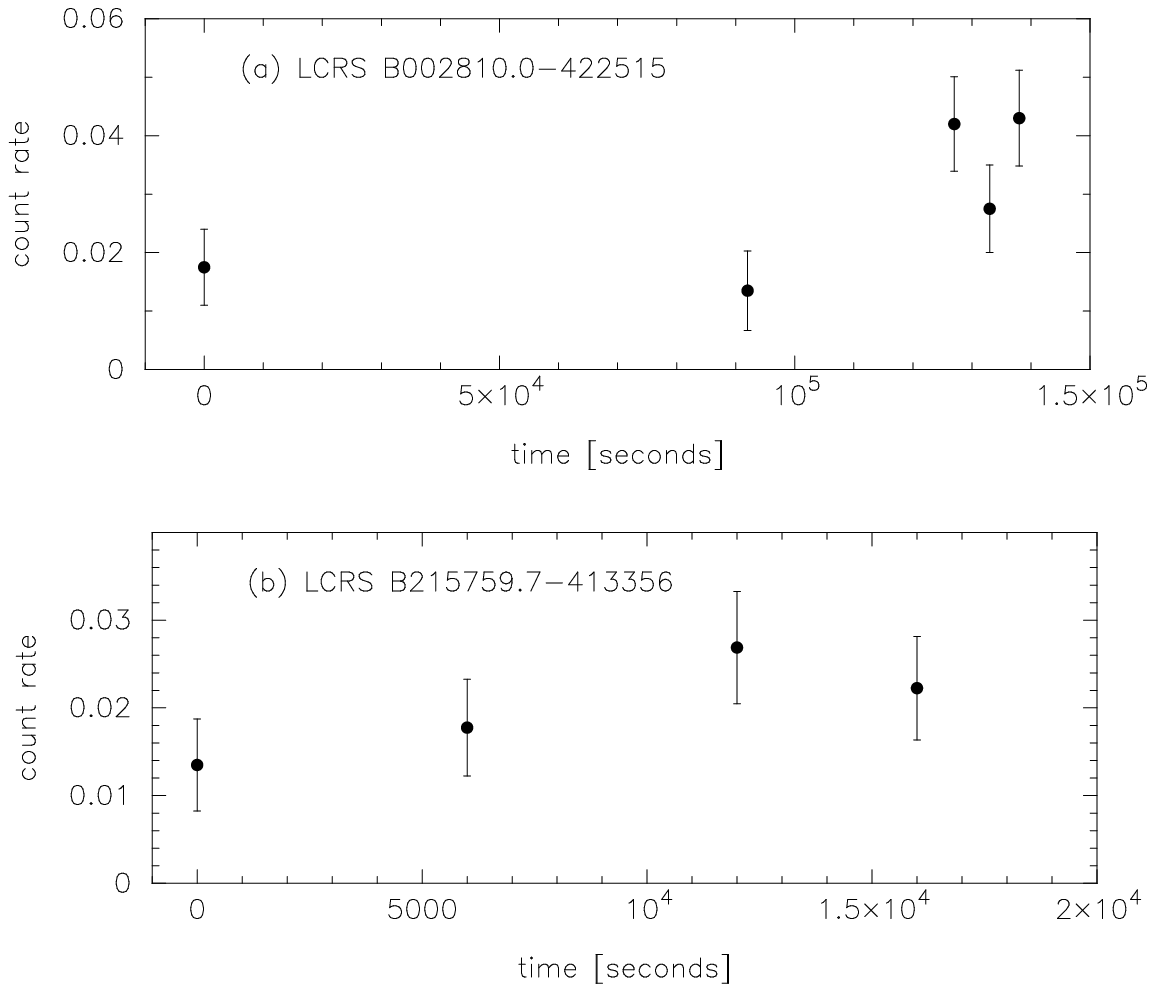}
\vspace{-10mm}
\aba
Fig.~4.  X-ray light curves for (a) LCRS~B002810.0-422515 and
(b) LCRS~B215759.7-413356.
\abe
\end{figure}

\end{document}